\documentclass[journal=jpcafh,manuscript=article]{achemso}
\setkeys{acs}{articletitle = true}
\usepackage{graphicx}
\usepackage{color}
\usepackage{subfig}
\usepackage{epstopdf}
\usepackage[utf8]{inputenc}

\usepackage{caption,float,geometry,natbib,setspace,xkeyval}

\title{Vibrational relaxation of hot ground state cations of naphthalene}

\author{Geert Reitsma}
\author{Serguei Patchkovskii}
\author{Judith Dura}
\affiliation{Max-Born-Institut f\"ur Nichtlineare Optik und Kurzzeitspektroskopie, Max-Born-Stra{\ss}e 2A, D-12489 Berlin, Germany}
\author{Lorenz Drescher}
\affiliation{Max-Born-Institut f\"ur Nichtlineare Optik und Kurzzeitspektroskopie, Max-Born-Stra{\ss}e 2A, D-12489 Berlin, Germany}
\altaffiliation{Present address: Department of Chemistry, University of California, Berkeley, California 94720, USA}
\author{Jochen Mikosch}
\author{Marc J. J. Vrakking}
\author{Oleg Kornilov}
\affiliation{Max-Born-Institut f\"ur Nichtlineare Optik und Kurzzeitspektroskopie, Max-Born-Stra{\ss}e 2A, D-12489 Berlin, Germany}
\email{oleg.kornilov@mbi-berlin.de}

\date{\today}

\begin{document}


\begin{abstract}
Time-resolved XUV-IR photoion mass spectroscopy of naphthalene conducted with broadband, as well as with wavelength-selected narrowband XUV pulses reveals a rising probability of fragmentation characterized by a lifetime of $92\pm4$~fs. This lifetime is independent of the XUV excitation wavelength and is the same for all low appearance energy fragments recorded in the experiment. Analysis of the experimental data in conjunction with a statistical multi-state vibronic model suggests that the experimental signals track vibrational energy redistribution on the potential energy surface of the ground state cation. In particular, populations of the out-of-plane ring twist and the out-of-plane wave bending modes could be responsible for opening new IR absorption channels leading to enhanced fragmentation.
\end{abstract}

\maketitle

\section{Introduction}
Non-adiabatic relaxation of aromatic molecular cations has been investigated in a number of recent XUV-pump-IR-probe experiments. Time-dependent dication yields in naphthalene and several other polyaromatic hydrocarbon (PAH) molecules were measured allowing to capture the relaxation of highly excited states just below the double ionization threshold \cite{marciniak_xuv_2015, reitsma_2019, herve_2021}. These studies indicate that relaxation timescales increase with the size of the molecule and with the excitation energy, which was also supported by a time-resolved photoelectron study of naphthalene \cite{marciniak_2019}. The XUV-IR  spectroscopy was also applied to benzene with improved temporal resolution by Galbraith et al \cite{galbraith_xuv-induced_2017}. The dynamics of the first six cationic states in benzene were investigated in detail by utilizing differences in the appearance energies of specific photoion fragmentation channels which could be activated by resonant IR photoabsorption in cations produced by the XUV \cite{galbraith_few-femtoseconds_2017}. 

Common to all these studies is the focus on relaxation of electronically excited states. In this work we  show that dynamics on the cationic ground state potential energy surface can be captured by XUV-IR photoion studies. We extend our recent investigation of naphthalene cations \cite{reitsma_2019}, using both broadband as well as wavelength-selected XUV pulses and monitor the production of fragment ions as a function of XUV-IR delay. The ionization potential (IP) of naphthalene is 8.12~eV \cite{jochims_photofragmentation_1992}. The appearance energy (AE) of the first fragmentation channel $\rm{C_{10}H_{8}^+}\longrightarrow\rm{C_{10}H_{7}^+}+\rm{H}$ is 15.35 eV, while the $\rm{C_2H_2}$, $\rm{2H/H_2}$, $\rm{C_4H_2}$, and $\rm{C_3H_3}$-loss channels have similar AEs of 15.36, 15.60, 15.65, and 16.09 eV, respectively \cite{jochims_photofragmentation_1992}. Multi-step fragmentation processes have significantly higher AEs, starting at 18.5 eV \cite{ruhl_single_1989,jochims_photofragmentation_1992}. Therefore channels with low AEs dominate the XUV-only spectra, in accordance with previous photodissociation studies \cite{west_dissociation_2012}. Our results demonstrate, that upon absorption of an extra IR photon hot ground state cations also fragment mostly into these channels. Therefore, the $\rm{H}$-, $\rm{C_2H_2}$-, $\rm{2H/H_2}$-, $\rm{C_2H_4}$-, and $\rm{C_3H_3}$-loss channels can serve as the observable for intra-molecular vibrational energy redistribution (IVR) processes in the electronic ground state. 

\section{Experiment}
The experiments reported here were carried out using two experimental arrangements with broadband and wavelength-selected XUV pulses, respectively, which were described in Refs. \cite{galbraith_xuv-induced_2017,galbraith_few-femtoseconds_2017} and Ref. \cite{reitsma_2019}
In the case of experiments with broadband XUV pulses, the output of a Ti:Sa laser system (central wavelength 795~nm, pulse duration 25~fs, pulse energy 3~mJ per pulse, repetition rate 1~kHz) is spectrally broadened and temporally compressed by sending it through a gas-filled hollow core optical fiber. The fiber is 1~m long, has a diameter of 300~$\mu$m, and is statically filled with 1.5~bar of Ne gas. The pulses are compressed by a chirped mirror compressor and characterized using the SEA-F-SPIDER technique \cite{witting_characterization_2011}. The beam is then split into two parts, one to generate high harmonics and one serving as a probe IR beam. Both arms are equipped with movable wedges, in order to optimize the compression of both pulses. XUV pulses were generated in Xe gas and the spectrum was filtered using a 300~nm Al filter, which blocks IR light and XUV light with photon energies below 15~eV and above 72~eV. In the second experimental arrangement the output of the same Ti:Sa laser is used without further compression. One part is used to generate high harmonics in Kr gas and the second part is used as IR probe beam. A Time Delay Compensating Monochromator (TDCM) is used to select a harmonic with a particular XUV wavelength \cite{eckstein_dynamics_2015, reitsma_2019}. 

In both arrangements the beams are recombined and focused into the interaction region of a velocity-map-imaging (VMI) spectrometer. The VMI spectrometer is equipped with a repeller-integrated oven ensuring sufficient target density and a well-defined interaction region. The VMI is mainly operated in a time-of-flight (TOF) mode. However, the VMI mode is used for reference experiments, such as independent measurements of the XUV-IR cross-correlation using Ar sidebands, which yielded an XUV-IR cross-correlation of 8.6~fs in case of the broadband source and 35-40~fs for the TDCM setup. The compression of the IR probe pulse was optimized by enhancing above threshold ionization (ATI) in Ar. The XUV pulse was optimized by enhancing the intensity of the XUV beam. The spectral resolution of the TDCM setup was 300 meV. The IR intensity was measured using the ponderomotive shift of photoionization lines in Ar when the XUV and the IR pulse were overlapped ($t=0$).

\begin{figure*}
\begin{center}
\subfloat[$\rm{C_{10}H_8^{2+}}$]{\includegraphics[width=7cm]{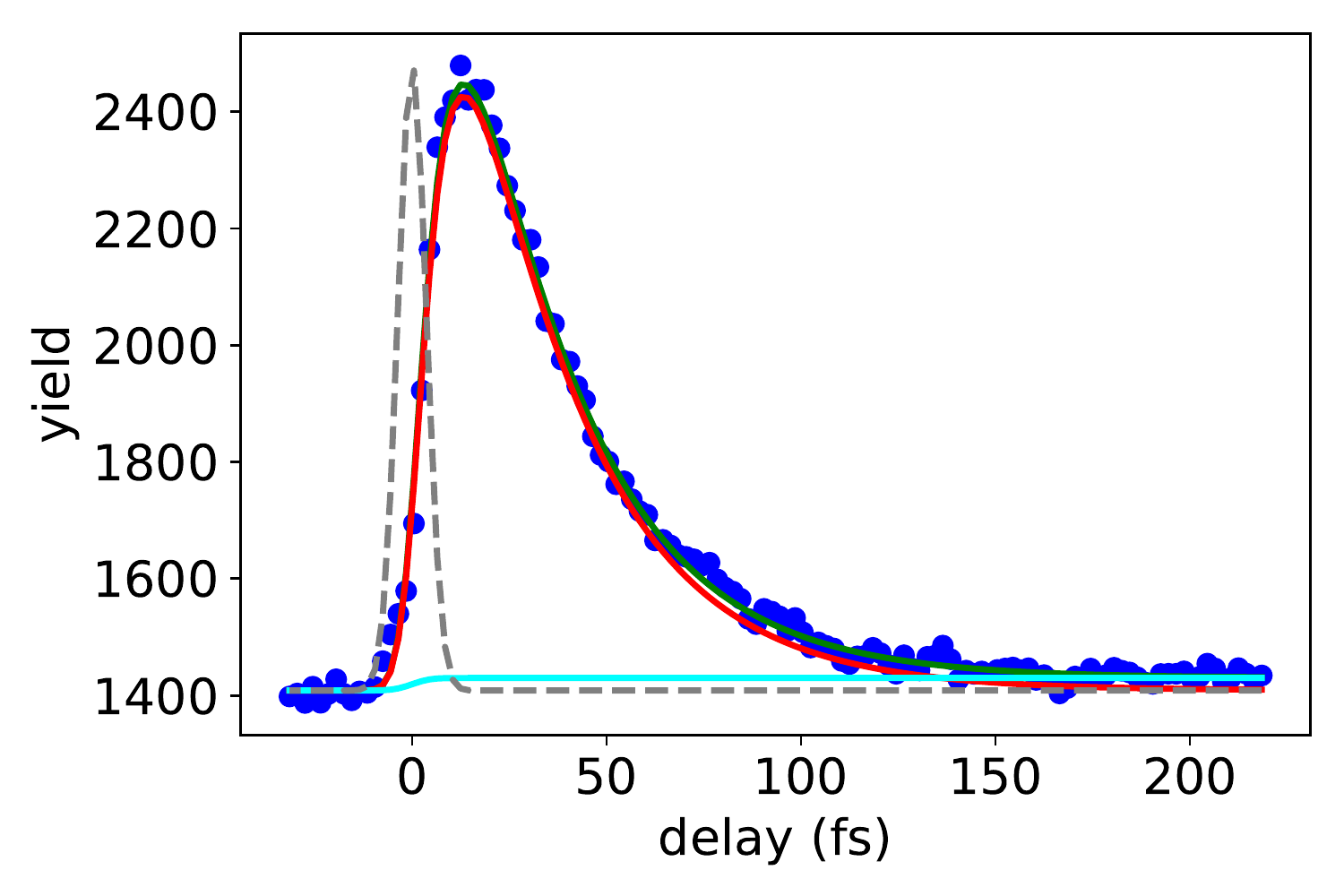}}\qquad
\subfloat[$\rm{C_6H_6^+}$ ($\rm{C_4H_2}$-loss)]{\includegraphics[width=7cm]{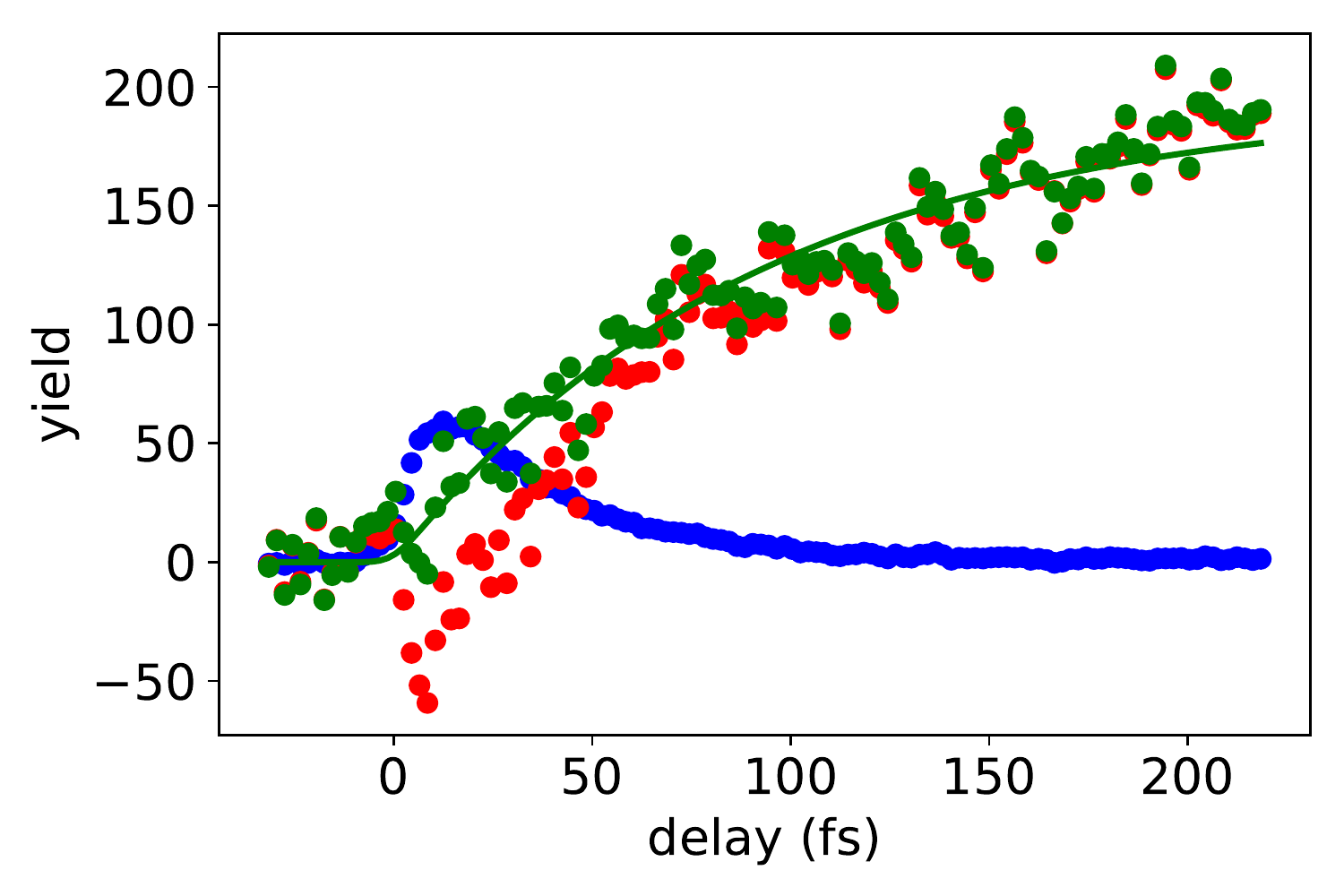}}\\
\subfloat[$\rm{C_8H_6^+}$ ($\rm{C_2H_2}$-loss)]{\includegraphics[width=7cm]{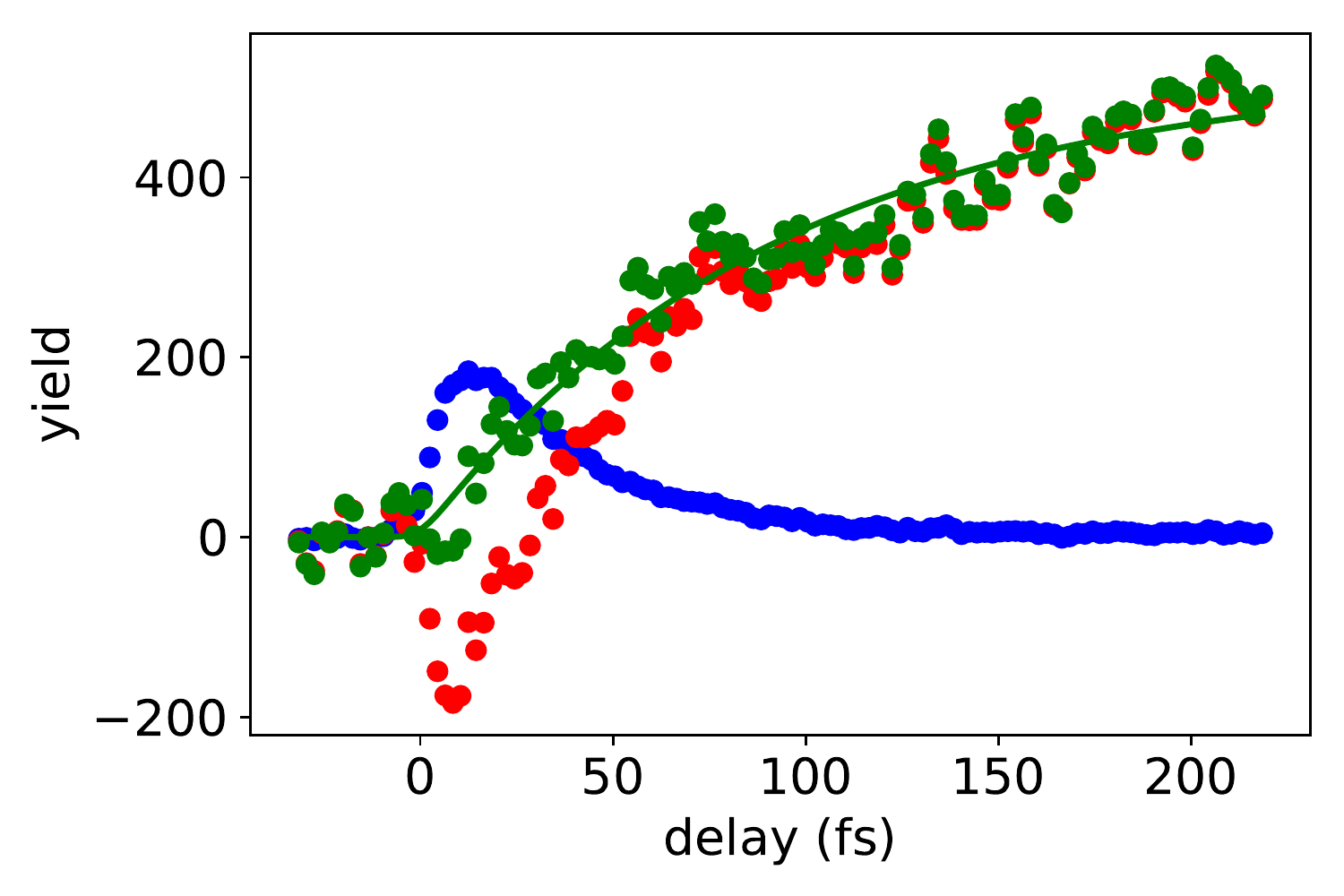}}\qquad
\subfloat[$\rm{C_{10}H_7^+}$ ($\rm{H}$-loss)]{\includegraphics[width=7cm]{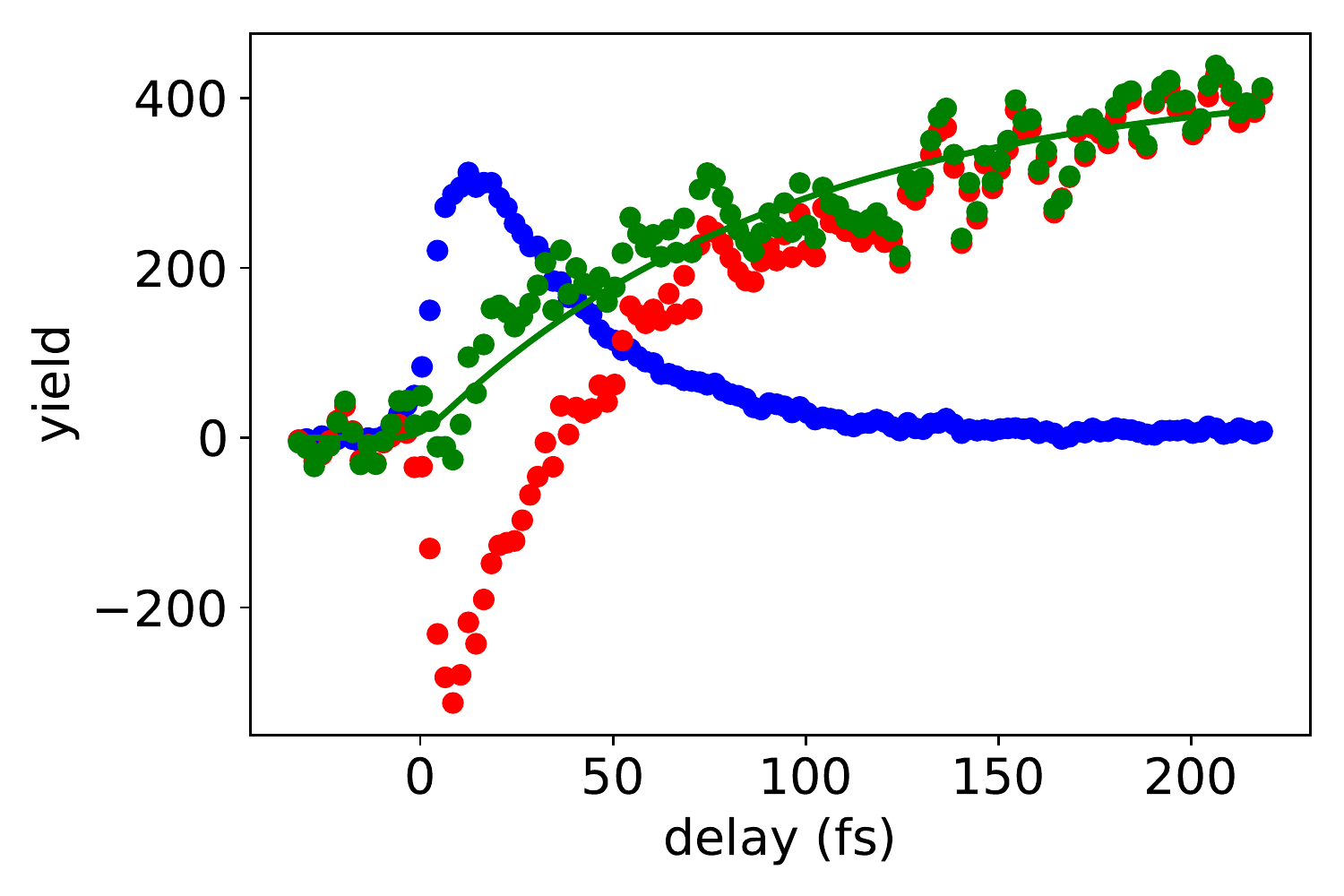}}\\
\caption{(a) Time-resolved yield of the dication $\rm{C_{10}H_8^{2+}}$ in two-color XUV-IR dissociative photoionization of naphthalene $\rm{C_{10}H_8}$, obtained using the broadband XUV source. The black dashed line represents the independently determined XUV-IR cross correlation of 8.6~fs. The blue dots are the experimental data. The red line is a fit to the dynamics of the dication, while the light blue line represents a long-lived component due to contamination from other channels (see text). Time-resolved yields of (b) $\rm{C_6H_6^+}$ ($\rm{C_4H_2}$-loss), (c) $\rm{C_8H_6^+}$ ($\rm{C_2H_2}$-loss) and (d) $\rm{C_{10}H_7^+}$ ($\rm{H}$-loss) fragments obtained in the same experiment. The red dots are the experimental data with the XUV-only sinal subtracted. The blue dots are the scaled dication signal from part (a). The green dots are the sum of these two components and the green line is the exponential rise fit to the green dots. See text for details of the data analysis and fitting.}
\label{strangedynamics}
\end{center}
\end{figure*}

\section{Results and Discussion}
Figure \ref{strangedynamics} shows the time-dependent yield of four photoions as a function of the XUV-IR delay obtained using the broadband source. A positive delay time means that the IR pulse arrives after the XUV pulse. Part (a) of the figure shows the transient for the naphthalene dication ($\rm{C_{10}H_8^{2+}}$). As previously reported \cite{marciniak_xuv_2015, reitsma_2019, marciniak_2019}, around $t=0$ the yield of the dication increases and then decays exponentially. The small long-lived component after $t=0$ is due to a slight contamination of the measurement by cation fragmentation channels due to insufficient mass resolution. To extract the relaxation timescale, the transient was fitted using the following model:
\begin{equation}
F_1(t)=e^-{^{\frac{(t-t_0)^2}{2\sigma^2}}} \ast \left(H(t-t_0)\left(A(e^{\frac{-(t-t_0)}{\tau_1}}-e^{\frac{-(t-t_0)}{\tau_2}})+B_1\right)\right)+B_2, 
\label{f1}
\end{equation}
where $2\sqrt{2\ln{2}}\sigma$ is the Full Width at Half Maximum (FWHM) of the XUV-IR cross correlation and $t_0$ is the time at which both pulses are temporally overlapped. The time constants $\tau_1$ and $\tau_2$ represent a rise time and a decay time of the transient. $H(t-t_0)$ is the Heaviside function and $*$ designates the convolution of the molecular response with a Gaussian profile representing the XUV-IR cross-correlation of the experiment. 

The fast dynamics observed in the dication track the electron-correlation driven population of the highly excited states of the naphthalene cation \cite{marciniak_xuv_2015,reitsma_2019}. These states are located just below the double ionization threshold and only one or two IR photons are sufficient to further ionize the molecule. The decay time of $\tau_1=30\pm1$~fs extracted from the fit is in the good agreement with the timescale that was reported previously \cite{marciniak_xuv_2015}. The rise time $\tau_2=7\pm1$~fs has not been reported previously. We note that this timescale represents delayed rise of the dication signal. A similar delay has previously been observed in benzene \cite{galbraith_xuv-induced_2017} and phenylalanine \cite{calegari_ultrafast_2014}. Galbraith and co-authors elaborated that this timescale may correspond either to ultrafast electronic relaxation of the initially excited  state or to dynamics in the Frank-Condon region.

Figure \ref{strangedynamics}(b-d) (red dots) displays the time-dependent yields of the three most abundant fragments with low appearance energies: $\rm{C_6H_6^+}$, $\rm{C_8H_6^+}$, and $\rm{C_{10}H_7^+}$,  corresponding to the emission of neutral $\rm{C_4H_2}$, $\rm{C_2H_2}$, and H, respectively. The three transients (b-d) show qualitatively similar behavior, which is very different from that of the dication (Figure \ref{strangedynamics}(a)). Near zero delay, corresponding to overlap of the XUV and IR pulses, the signal is depleted, but it then quickly recovers and continues to grow. This indicates that production of these low AE fragments is enhanced by the IR pulse, when it follows the XUV excitation.

To analyze the observed time-dependent dynamics we first note, that the depletion at small delays is expected to reflect the enhancement of the dication signal. When the XUV photon populates high-lying states, the molecule rapidly relaxes to the ground state and/or lower excited states with the electronic excitation energy transferred to vibrational degrees of freedom \cite{reitsma_2019}. This stored energy eventually leads to molecular fragmentation, mainly to the channels with low AE. However, when the IR pulse further ionizes the cation before the non-adiabatic relaxation occurs, a part of the population is transferred to the dication, which has a different fragmentation pattern. This leads to a depletion of the $\rm{C_6H_6^+}$, $\rm{C_8H_6^+}$, and $\rm{C_{10}H_7^+}$ signals. Taking this into account we correct the experimental data in Figure \ref{strangedynamics}(b-d) (red dots) by adding the dication yield from the part (a) scaled such as to compensate for the observed depletion. We fit the resulting sum (green dots) to a rising component described by an exponential function (green line):
\begin{equation}
F_2(t)=e^-{^{\frac{(t-t_0)^2}{2\sigma^2}}} \ast \left(H(t-t_0)\left(C(1-e^{\frac{-(t-t_0)}{\tau_3}})\right)\right), 
\label{f2}
\end{equation}
where $C$ is the asymptotic yield of the fragment and $\tau_3$ is the timescale of the exponential rise. To improve the error of the fit we constructed a multi-dimensional fit model in which we fixed $\sigma$ and $t_0$ to the values obtained from the dication signal and defined the time-constant $\tau_3$ as a global parameter, i.e. fitting the rise time for all three fragments. This fitting procedure results in a time constant of $\tau_3=92\pm4$~fs.

We proceed to discuss the origin of the slow rise component corresponding to enhanced fragmentation induced by the IR pulse. Since the transient dynamics appears to be the same for all low AE fragments, we propose that also after the interaction with the IR pulse the fragmentation is of statistical nature and proceeds on nanosecond timescales from the hot ground state of the naphthalene cations prepared by the combined XUV+IR ionization. The action of the IR pulse may thus either be i) to increase the number of hot cations by ionizing highly excited neutral molecules or ii) to increase the temperature of the cations prepared by the XUV, leading to an enhancement of the fragmentation. To distinguish between these two scenarios we employed the second experimental arrangement (see the Experimental section) using the wavelength-selected XUV pulses. Controlled tuning of the XUV wavelength over the range covered by the broadband source helps to distinguish between the involvement of neutral and cationic species, because neutral excited molecules can only be prepared when the XUV wavelength is resonant with an existing transition. In the wavelength-dependent experiment dynamics due to neutral molecules would only be expected for certain choices of the XUV wavelength, while the cationic dynamics would be expected, independent of the choice of the photon energy.

\begin{figure*}
\begin{center}
\includegraphics[width=18cm]{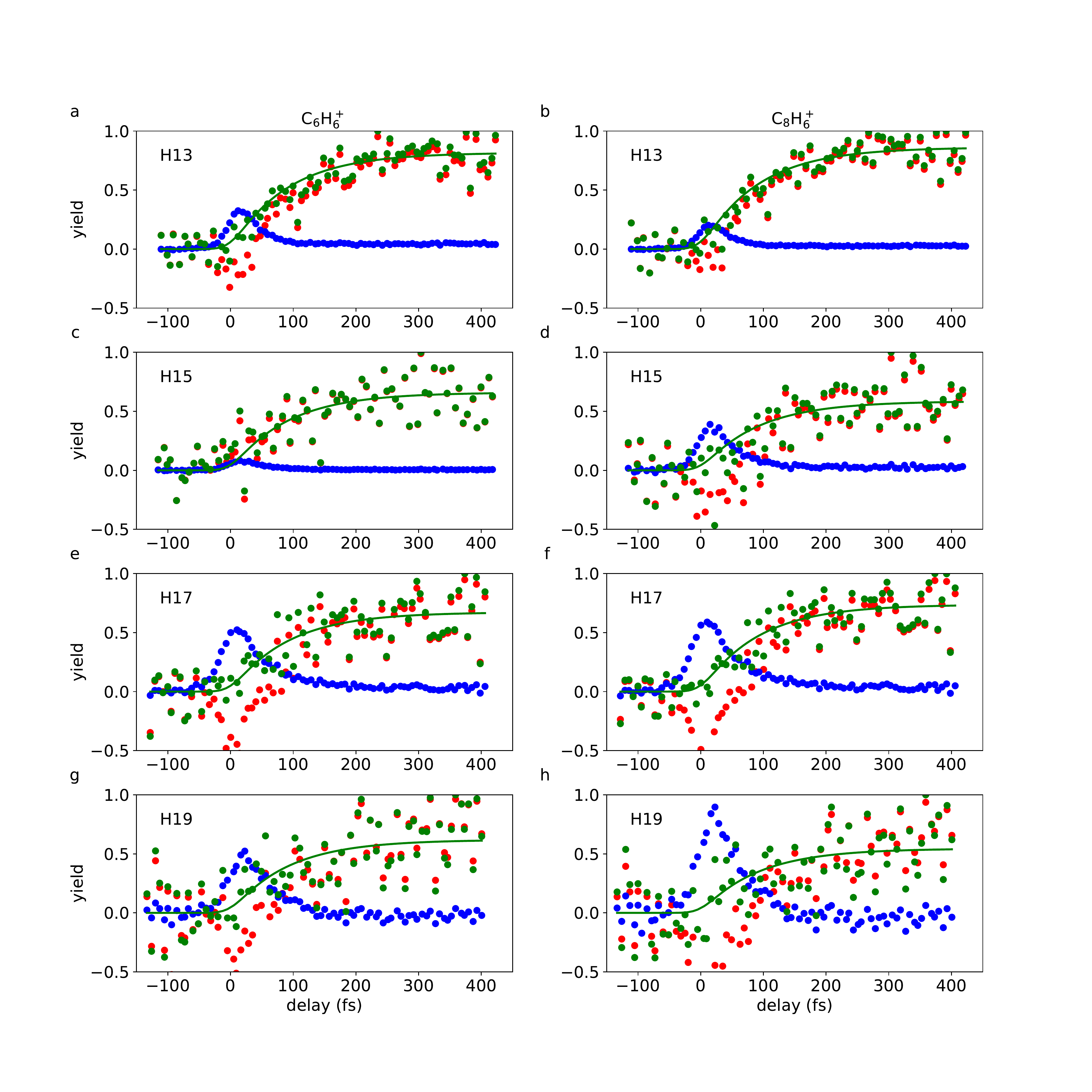}
\caption{Time-resolved yields of the (a,c,e,g) $\rm{C_6H_6^+}$ ($\rm{C_4H_2}$-loss) and 
		(b,d,f,h) $\rm{C_8H_6^+}$ ($\rm{C_2H_2}$-loss) fragments recorded with the wavelength-selected XUV pulses using odd harmonics 13 to 19 (20.3, 23.4, 26.5 and 29.6~eV). Red dots correspond to the measured yields, blue dots correspond to the scaled yield of the $\rm{C_{10}H_8^{2+}}$ dication measured in the same experimental run. Green dots are the sum of the two yields with appropriate scaling coefficients and green lines are exponential rise fits. See text for details. 
}
\label{mono}
\end{center}
\end{figure*}

Figure \ref{mono} displays time-dependent yields of the $\rm{C_6H_6^+}$ ($\rm{C_4H_2}$-loss) and $\rm{C_8H_6^+}$ ($\rm{C_2H_2}$-loss) fragments obtained using odd XUV harmonics 13-19, which correspond to central photon energies of 20.3, 23.4, 26.5 and 29.6~eV, respectively. Note that the H-loss channel was insufficiently mass-resolved and therefore excluded from the analysis. Similar to the  Figure \ref{strangedynamics}(b-d), the red dots represent the yields of (a) $\rm{C_6H_6^+}$ and (b) $\rm{C_8H_6^+}$ fragments, the blue dots represent the scaled time-dependent dication yield for each harmonic, and the green dots represnt the sum of these two components. Once more, the exponential rise model (green line) with the fixed global rise time of $\tau_3=92$~fs (taken from the first experiment) very well describes the transients for both fragments and for all harmonics. We thus conclude that a significant contribution from neutral species can be excluded. The $\tau_3$ timescale should thus be assigned to relaxation of naphthalene cations towards molecular states with an increased IR absorption cross section. The energy deposited by the IR pulse increases the number of cations with temperature above the dissociation threshold and thus increases the fragmentation yields. As possible states responsible for the absorption one can consider i) low-lying electronically excited states which open resonant IR absorption channels or ii) the vibrationally excited ground state. 

The theoretical potential energy surfaces of the lowest six cationic states in
naphthalene and the dynamics on these surfaces have been theoretically studied by Ghanta
et al \cite{ghanta_theoretical_2011,ghanta_theoretical_2011-1}. Within the MCTDH approach the
time-dependent populations of the $\rm{\widetilde{X}}$, $\rm{\widetilde{A}}$,
$\rm{\widetilde{B}}$, $\rm{\widetilde{C}}$, $\rm{\widetilde{D}}$, and
$\rm{\widetilde{E}}$ were calculated upon initial population of the
$\rm{\widetilde{A}}$, $\rm{\widetilde{B}}$, $\rm{\widetilde{C}}$,
$\rm{\widetilde{D}}$, or $\rm{\widetilde{E}}$ states. In most cases the
calculated decay timescales were much shorter than the timescale of $\approx100$~fs
observed in our experiments (see Figure 3 of Ref.
\cite{ghanta_theoretical_2011-1}). The only exception is transfer of the
population between the $\rm{\widetilde{B}}$ and $\rm{\widetilde{A}}$ states,
which proceeds on timescale of about 200~fs, i.e. slower than observed
here. However, we note that the $\rm{\widetilde{B}}$ band is located at 10.3~eV,
which is 5~eV below the lowest fragment appearance energy. The absorption
of three and more IR photons (1.56~eV) would be then required to enhance
fragmentation. Therefore we consider the assignment of the observed timescale
$\tau_3$ to relaxation processes in the manifold of electronically excited
states as improbable. 

Consequently, we now turn to the analysis of the second proposition, namely that the observed
transients reflect dynamics on the potential energy surface of the ground
electronic state after the electronic relaxation has largerly been completed. The photoabsorption spectrum for $\rm{C_{10}H_8^+}$ cations
has been studied both experimentally \cite{salama_electronic_1991} and
theoretically \cite{niederalt_ab_1995}. The most prominent feature is
the 0-0 transition from the electronic ground state
$\rm{\widetilde{X}^2A_u(D_0)}$ to the second excited state
$\rm{^2\widetilde{B}_{3g}(D_2)}$ at 675~nm \cite{salama_electronic_1991}. No
features were identified in cold cations at or close to 795~nm - the probe
wavelength in our experiments. The absorption spectrum may change however, when the
cations are highly vibrationally excited, as expected when the
electronic energy of the excited cations is converted to the vibrational energy
of the ground state cations during relaxation via a series of conical
intersections.

\begin{figure*}
\begin{center}
\includegraphics[width=18cm]{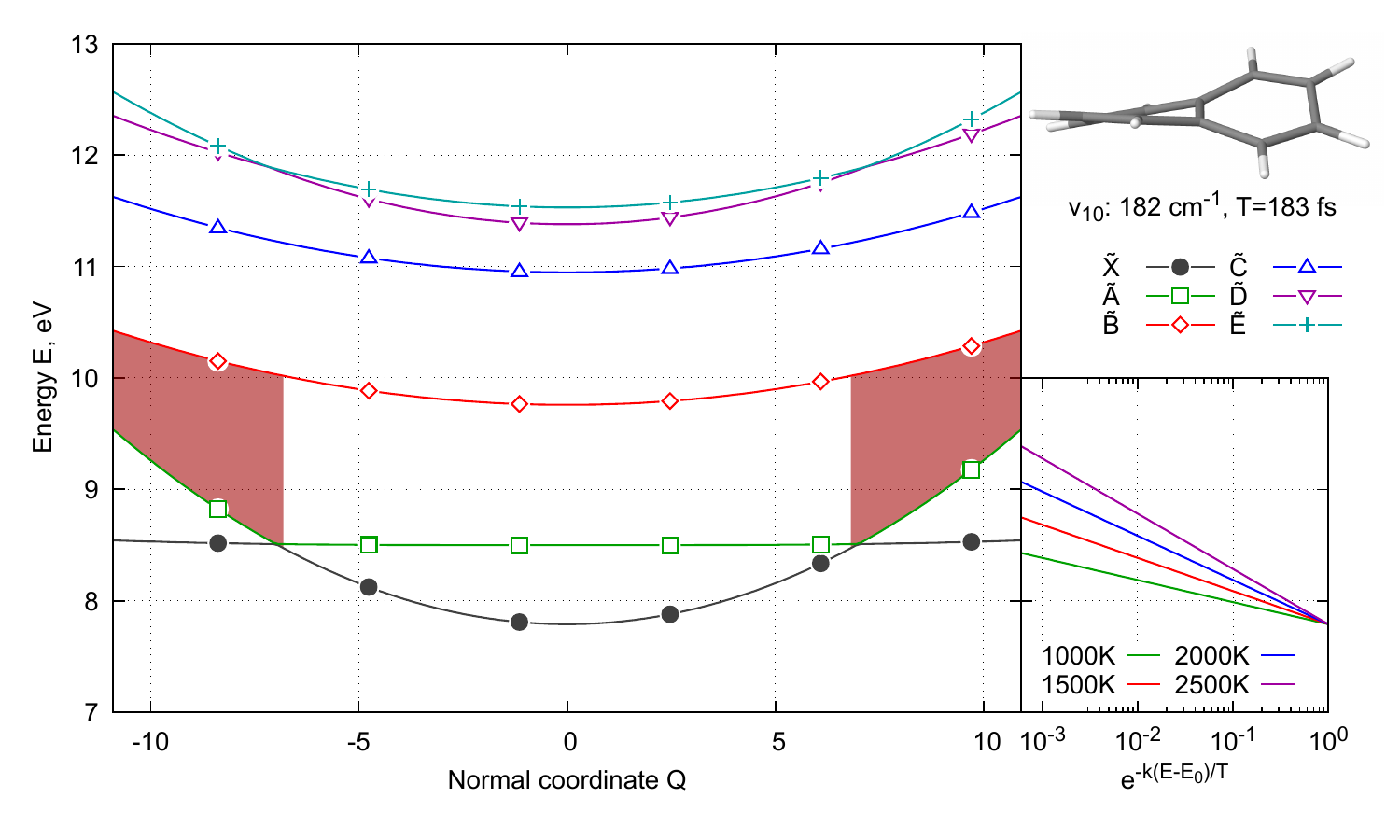}
\caption{Cut through the potential energy surfaces of the naphthalene cation along
the $\nu_{10}$ normal mode\cite{ghanta_theoretical_2011}. The adiabatic
electronic states are denoted by the corresponding diabatic state label at the
neutral equilibrium geometry ($Q=0$). Note, that adiabatic states do not cross, but can touch each other due to high degree of degeneracy. The shaded areas indicate ranges of $Q$ where  dipole transitions are accessible for the IR photon energy used in the present experiment (1.56~eV). The right panel indicates the Boltzmann factors calculated for temperatures between $1000$ and $2500$~K. The vertical energy axis is aligned with the axis of the main plot and energies are taken relative to the minimum of the $\tilde X$ cationic ground state ($E_0=7.79$~eV). The insert in
the top right corner illustrates the structural distortion along the $\nu_{10}$
($a_{2u}$) normal mode.
}
\label{pes_nu10}
\end{center}
\end{figure*}

\begin{figure*}
\begin{center}
\includegraphics[width=18cm]{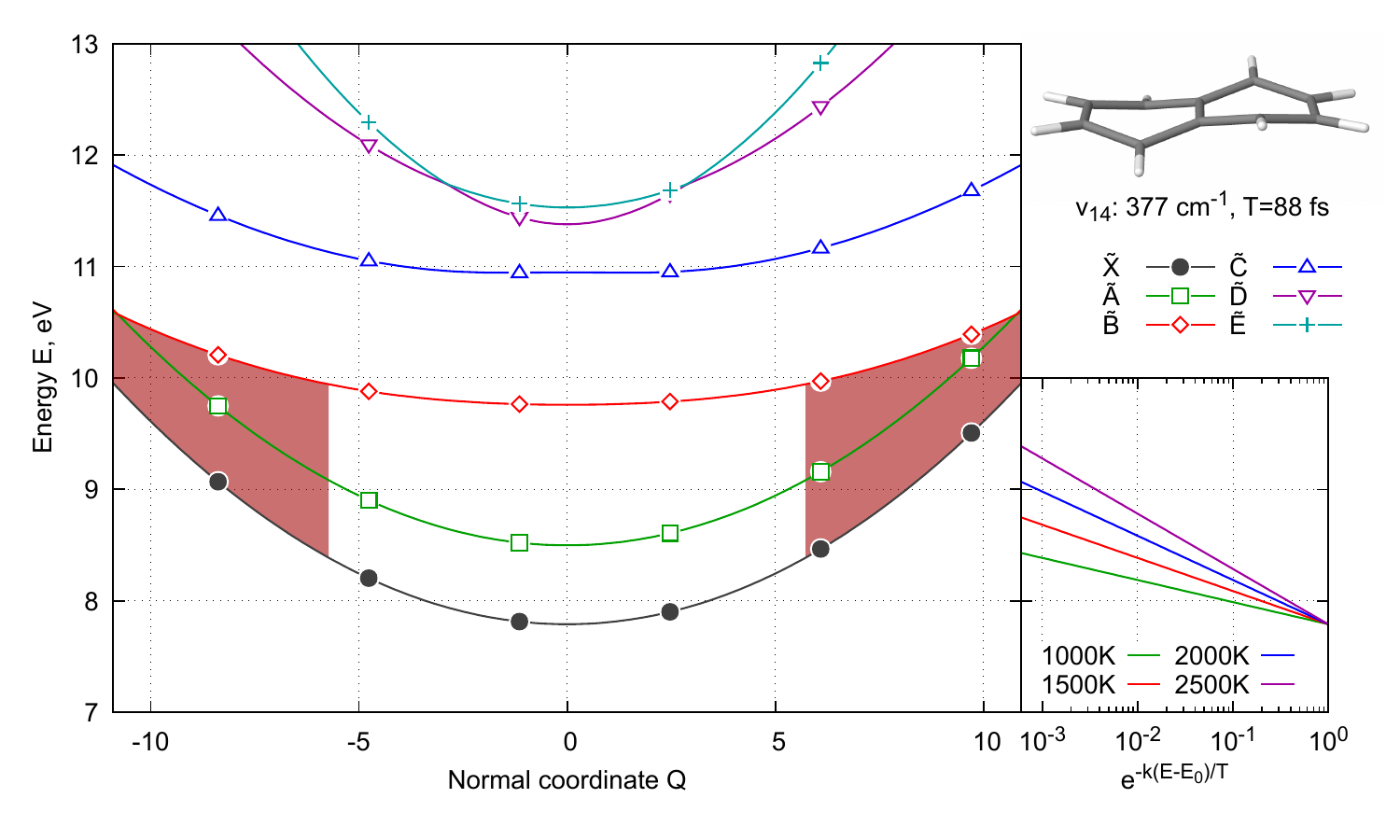}
\caption{Cut through the potential energy surfaces of the naphthalene cation along
the $\nu_{14}$ ($b_{1g}$) normal mode\cite{ghanta_theoretical_2011}. See 
caption of Figure~\ref{pes_nu10}.
}
\label{pes_nu14}
\end{center}
\end{figure*}

To examine this assumption we constructed a model based on the vibronic
Hamiltonian of Ghanta et al \cite{ghanta_theoretical_2011}. The Hamiltonian
contains all inter-mode and inter-state couplings for the six lowest diabatic
states of the naphthalene cation. We approximate the transition dipoles between
the diabatic states by the dipole matrix elements between Hartree-Fock
molecular orbitals of neutral naphthalene (cc-pVDZ basis set) at the optimized
MP2/cc-pVDZ geometry (the $Q=0$ point of Ghanta et al
\cite{ghanta_theoretical_2011}). In this approximation, the $\tilde X^2A_u$
diabatic state is strongly coupled to the $\tilde B^2B_{2g}$ ($d=6.6$~Debye)
and $\tilde C^2B_{1g}$ ($d=4.8$~Debye) electronic states. The $\tilde C$ state
is in turn dipole-coupled to the $\tilde A^2B_{3u}$ state ($d=7.6$~Debye),
while the $\tilde B$ state has no further strong dipole transitions.

We explored the temperature dependence of the IR absorption separately for each
normal vibrational mode of the vibronic Hamiltonian, without considering combination modes. It is assumed that for each normal mode, the relative populations of the six
adiabatic electronic states follow classical Boltzmann statistics. At the high
temperatures considered here (see below), quantum vibrational effects are
not expected to be important, and are neglected. The probability of an
IR-induced transition between all pairs of adiabatic electronic states is
determined from Fermi's golden rule, for a random molecular orientation. 

The effective temperature range for our calculation is rationalized as follows.
The XUV photon energies used in our experiment are sufficiently high to produce
highly excited naphthalene cations both below and above the appearance
energy of the $\rm{C_6H_6^+}$ ($\rm{C_4H_2}$-loss) and $\rm{C_8H_6^+}$
($\rm{C_2H_2}$-loss) fragments. Given the naphthalene ionization potential of 8.12~eV
\cite{tobita_single-ionization_1994}, population of states just below the
cation dissociation threshold of 15.35~eV leads to the internal energy of
7.23~eV. If we assume an equipartitioning of this energy over all vibrational
degrees of freedom of the molecule, we arrive at an energy of $7.23/(3*18-6)=0.15$~eV per mode, which corresponds to a cation temperature of 1700~K.  Again
assuming energy equipartitioning, the absorption of each IR photon further
increases this temperature by about 370~K.

A detailed analysis of the results shows that only excitations of 8 out of the 48
vibrational modes of naphthalene influence the IR-induced transition rate
at temperatures below $2500$~K. The dominant contribution is given by $\nu_{10}$
(the out-of-plane ring twist) and $\nu_{14}$ (the out-of-plane wave bend).

A cut through the adiabatic potential energy surfaces along the $\nu_{10}$
normal mode is shown in Figure~\ref{pes_nu10}. The two low-lying adiabatic
electronic states (correlating to $\tilde X^2A_{u}$ and $\tilde A^2B_{3u}$ at
$Q=0$) are significantly populated for temperatures between $500$ and $2500$~K
as indicated in the inset showing Boltzmann factors for the corresponding
energies. Both states are dipole-coupled to the $\tilde B^2B_{2g}$ (red), in
the region of $Q$ parameters where the diabatic $\tilde X^2A_{u}$ state has a
dominant contribution. Both states are also dipole-coupled to the $\tilde C^2B_{1g}$
(blue) electronic state; however, only the $\tilde B$ state can be
accessed with a single IR photon. The calculated rate of the IR absorption due
to the $\nu_{10}$ motion follows the Arrhenius' law
[$k\propto\exp\left(-T_a/T\right)$].  Linear fit to the calculated rates for
$T=250$--$2250$~K in the $T^{-1}$--$\log{k}$ coordinates yield activation
temperature $T_a\approx 7.2\times10^3$~K. The harmonic vibrational frequency
for the $\nu_{10}$ mode is $182$($399$)~cm$^{-1}$ [calculated
(experimental)]\cite{ghanta_theoretical_2011}. The corresponding vibrational
period is $183$($84$)~fs. Both values are compatible with the experimental time
constant we determine ($\tau_3=92\pm4$~fs).

A similar analysis for the $\nu_{14}$ normal mode is presented in
Figure~\ref{pes_nu14}. Again, the thermally accessible $\tilde X$ and $\tilde A$
states can be brought to resonance with the dipole-accessible $\tilde B$ state.
The second dipole-allowed transition, to the $\tilde C$ state, may also become
accessible at higher temperatures. The Arrhenius activation temperature is now
$T_a\approx6.5\times10^3$~K. The harmonic vibrational frequency for the $\nu_{14}$ mode 
is $377$($191$)~cm$^{-1}$\cite{ghanta_theoretical_2011}, giving a vibrational period
of $88$($175$)~fs. Again, these values are compatible with the experimentally determined
timescale. 

In the absence of a detailed quantum-dynamics study, it does not appear to be
possible to differentiate between the $\nu_{10}$ and $\nu_{14}$ out-of-plane
vibrational modes as the primary vehicle for the IR-assisted dissociation.
In reality, both are likely active.

The discussion above allows us to interpret the observed  slow rise in the two-color
yield of low AE fragments as intra-molecular vibrational energy redistribution (IVR). Upon ionization the manifold of highly-excited electronic
states of the cation is populated. These states are strongly coupled to the
ground state of the cation via a series of conical intersections. During and after electronic relaxation the excitation energy flows into various vibrational modes. The fastest rate at which the energy can flow into a mode can be estimated from its eigenfrequency. As the populations of the two relevant modes, the $\nu_{10}$ and $\nu_{14}$ modes, increases the possibility of IR absorption opens up (see the shaded areas in Figures~\ref{pes_nu10} and \ref{pes_nu14}), leading to further energy deposition into the molecule and to the enhancement of fragmentation. 

\section{Conclusions}

In this work we investigated relaxation dynamics of naphthalene cations by means of time-resolved XUV-IR photofragment ion mass spectroscopy. We employed two complementary HHG-based XUV sources: a broadband source with laser pulses compressed using the hollow core fiber technique, which delivered a superior experimental time resolution of 8.6~fs; and a wavelength-selected source employing a time-delay-compensating monochromator, which delivers inferior time resolution of 35-40~fs, but provides a narrow and tunable XUV spectrum. The results of measurements in both arrangements broadly agree with each other and reveal a new exponential rise component in naphthalene cation fragmentation channels with low appearance energies, namely, the $\rm{C_6H_6^+}$ ($\rm{C_4H_2}$-loss), $\rm{C_8H_6^+}$ ($\rm{C_2H_2}$-loss) and  $\rm{C_{10}H_7^+}$ (H-loss) channels, in addition to the previously observed dynamics in the dication $\rm{C_{10}H_8^{2+}}$. In all experiments the same time constant of $92\pm4$~fs can consistently describe the exponential rise.

We analyzed possible origins of these dynamics. The fact that the same time scale is observed for many XUV excitation wavelengths and for all low AE fragments suggests that highly excited neutral and cationic states can be excluded from consideration and that the observed experimental signals are caused by dynamics on the hot ground electronic state of naphthalene cations. To gain further insight we analyzed the results based on the comprehensive vibronic Hamiltonian of Ghanta et al \cite{ghanta_theoretical_2011, ghanta_theoretical_2011-1}. We constructed a statistical model of vibrational excitations, which revealed that activity in the out-of-plane ring twisting mode ($\nu_{10}$) and the out-of-plane wave bending mode ($\nu_{14}$) may facilitate energy deposition in the molecule by the probe IR pulse. The vibrational periods of these modes are in good agreement with the observed timescale. 

\section*{Acknowledgements}
We thank Franck L\'epine for fruitful discussions and Ahmet \"Unal for experimental support. G.R. thanks the Netherlands Organization for Scientific Research (NWO) for financial support (Rubicon 68-50-1410). O. K. acknowledges support of the Deutsche Forschungsgeminschaft (KO 4920/1-1).

\bibliography{atto_naphth_4}
\end{document}